# Reversible control of Dzyaloshinskii-Moriya interaction at graphene/Co interface via hydrogen absorption

Baishun Yang[1], Qirui Cui[1], Jinghua Liang[1], Mairbek Chshiev[2*], Hongxin Yang[1,3*]

[1] *Ningbo Institute of Materials Technology and Engineering, Chinese Academy of Sciences, Ningbo 315201, China*

[2] *Univ. Grenoble Alpes, CEA, CNRS, Grenoble INP, IRIG-SPINTEC, Grenoble, France*

[3] *Center of Materials Science and Optoelectronics Engineering, University of Chinese Academy of Sciences, Beijing 100049, China*

*Email：mair.chshiev@cea.fr, hongxin.yang.spintec@gmail.com

ABSTRACT:

Using first-principles calculations, we investigate the impact of hydrogenation on the Dzyaloshinskii-Moriya interaction (DMI) at graphene/Co interface. We find that both the magnitude and chirality of DMI can be controlled via hydrogenation absorbed on graphene surface. Our analysis using density of states combined with first-order perturbation theory reveals that the spin splitting and the occupation of Co-*d* orbitals, especially the $d_{xz}$ and $d_{z^2}$ states, play a crucial role in defining the magnitude and the chirality of DMI. Moreover, we find that the DMI oscillates with a period of two atomic layers as a function of Co thickness what could be explained by analysis of out-of-plane of Co orbitals. Our work elucidates the underlying mechanisms of interfacial DMI origin and provides an alternative route of its control for spintronic applications.

Topological magnetic textures, such as magnetic skyrmions [1-6] and chiral domain walls [7, 8] can be used as information carriers for next generation information storage and logic technologies thanks to their high stability, small size and fast current driven mobility [9]. The Dzyaloshinskii-Moriya interaction (DMI) [10, 11], which originates from spin-orbit coupling (SOC) in inversion symmetry broken system, plays a crucial role in the formation of these topological magnetic textures. Specifically, it can influence the chirality as well as stability and migration velocity of chiral domain walls and skyrmions [6, 7]. Therefore, finding an efficient approach to control the magnitude and chirality of DMI is beneficial for creation and manipulation these magnetic textures for graphene spintronic applications.

Recent reports indicate that the DMI can be induced at graphene/ferromagnetic metal interface [12, 13]. Meanwhile, graphane, graphone and one-third-hydrogenated graphene representing different proportion of C:H (1:1, 2:1 and 3:1, respectively) are successfully fabricated experimentally with large scale [14-17]. These experimental results imply the possibility of changing the concentration of H at graphene/ferromagnetic metal surface. Moreover, the smallest size, the lowest atomic weight and very weak binding energy of hydrogen make it attractive for devices employing hydrogen migration. Indeed, the modulation of hydrogen can significantly influence the magnetism of materials as reported, for instance, by A. J. Tan *et al*. who achieved 90° magnetization switching with $H^+$ insertion at Co/GdO$_x$ interface by external electric field [18]. Furthermore, it was also reported that multiferroics with strong magnetoelectric coupling was realized in antiferromagnetic $SrCoO_{2.5}$ through hydrogen intercalation [19]. Both experiments and theories made progress in tuning the DMI by changing materials, atomic layer stacking, insulator capping and external electric field etc.[20-22]. However, a possibility of controlling the DMI via hydrogenation has not yet been reported.

In this Letter, using first-principles calculations, we systematically investigated the influence of DMI by varying the concentration of H at graphene surface of graphene/Co structures. We found that the DMI oscillates with a period of about two monolayers(ML) as a function of Co film thickness. More importantly, we demonstrated that not only the strength but also the chirality of DMI could be controlled by changing the concentration of hydrogen absorption.

The Dzyaloshinskii-Moriya interaction is calculated by employing the constrained spin-spiral supercell method [23]. Fig. 1(a) shows the schematic diagram of hydrogenated graphene/Co structures with clockwise (left panel) and anticlockwise (right panel) spin textures in one supercell. The ratio between H and C atom is varied from zero to half and to distinguish the concentration of hydrogen atom in the following, we label the five systems as Co@Gr, Co@Gr-1/8H, Co@Gr-2/8H, Co@Gr-3/8H, Co@Gr-4/8H, respectively. Please note that in following discussions on DMI sign and chirality we will always assume graphene on top of Co despite aforementioned notations used for convenience and will adopt positive (negative) DMI representing clockwise (anticlockwise) chirality. All of our calculations are performed within the framework of density functional theory (DFT) implemented in Vienna *ab initio* simulation package (VASP) [24-26]. The exchange-correlation potential is treated with the generalized gradient approximation (GGA) with Perdew-Burke-Ernzerhof (PBE) functional [27]. The cutoff energy is 520 eV and a 6×24×1 Gamma-centered *k*-mesh is used in the calculations. A vacuum region larger than 15 Å is adopted in all the calculations to avoid the interaction between the neighboring slabs. The atomic positions are fully relaxed until the force and total energy is less than 0.001 eV/Å and $10^{-7}$ eV, respectively.

The optimized Co(4ML)@Gr-4/8H slab is shown in Fig. 1(a). The distance between C and H is 1.13 Å which is nearly the same as that in graphane and graphone [28]. With the absorption of hydrogen, the C atom bonding with H atom moves out of the plane toward H by 0.44 Å which makes the graphene not planar any more, while the cobalt atoms are still in the same plane. Fig. 1(b) shows the calculated total DMI, $d^{tot}$, of graphene/Co with different concentration of H absorption when the thickness of Co varies from 1ML to 7ML. One can see that in all cases the DMI oscillates with a period of 2ML as a function of Co film thickness up to 5ML beyond which the DMI strength nearly stabilizes at constant values. Furthermore, the magnitude of DMI gradually increases as a function of hydrogen concentration and even changes its sign in some cases. For example, the DMI strength in Co(1ML)@Gr is -1.14 meV and it reverses its sign and decreases to 0.23 meV in Co(1ML)@Gr-4/8H.

Considering that the oscillatory DMI behavior of RKKY type is a common feature in all systems considered here, we suppose that this phenomenon is not due to hydrogen. The oscillatory behavior as a function of film thickness due to quantum well states (QWS) occurs for different physical phenomena such as magnetocrystalline anisotropy [29-31], Curie temperature [32] and magnetic exchange coupling [33-35]. In particular, similar oscillation period of 2ML due to QWS in Co film was reported for magnetic anisotropy energy variation as a function of Co thickness in fcc-Co slabs and $L1_0$-MnGa/Co(Fe) films [36, 37]. Quantum well states in *hcp*-Co case considered here originate from the electronic states reflected from top graphene/Co and bottom Co/vacuum interfaces. The DMI of graphene/Co interface is induced by the Rashba effect with the SOC energy localized on Co atoms [12]. In order to elucidate the origin of DMI oscillations, in Fig. 1 (c) we show the associated SOC energy dependencies as a function of Co thickness resolved for each $d$ orbital of Co atoms. It can be seen that for both Co@Gr and Co@Gr-4/8H systems, the associated SOC energy arising from $d_{xz}$ and $d_{z^2}$ has an oscillation period of 2ML with a large amplitude, so does of the $d_{yz}$ orbital but with a smaller amplitude. In contrast, the SOC energy contributions from in-plane $d_{xy}$ and $d_{x^2-y^2}$ orbitals are almost degenerated in respect to each other and have almost no change as a function of Co thickness. Thus, the oscillations of DMI is determined by the $d_{z^2(xz,yz)}$ states of Co film that could be attributed to much stronger impact of interfaces on out-of-plane orbitals compared to the in-plane ones.

Let us now investigate other physical mechanisms explaining the DMI variation as function of hydrogenation. A. Belabbes *et al*. [22] demonstrated that besides the SOC, the electronic occupation of magnetic atom also strongly affects the DMI. In particular, the DMI in 3$d$/5$d$ interfaces follows Hund's rule with a similar tendency to their magnetic moments [22]. To verify whether the underlying mechanism of the DMI change in our case is associated from the magnetic moment variation, we plot the corresponding average magnetic moment of each systems in Fig. 1(d). One can see that the average magnetic moment of Co decreases monotonically as hydrogen concentration increases with significantly more pronounced impact for thin Co films. For instance, the magnetic moment of Co drops from 1.65 $\mu_B$ in Co(1ML)@Gr to 1.00 $\mu_B$ in Co(1ML)@Gr-4/8H due to the strong interfacial hybridization between H and C. For large Co thicknesses, the magnetic moments stabilize within a narrow interval between 1.55 and 1.60 $\mu_B$. Interestingly, in most of the cases of thin Co films the average magnetic moment also shows oscillatory behavior with a period of 2ML. Most importantly, there is a clear correlation between magnetic moment and DMI curves with a higher concentration of hydrogen diminishes both the DMI and the magnetic moments. This suggests that the DMI also follows the Hund's rule here. However, the sign of DMI does not follow the trend of magnetic moment variation. Therefore, the Hund's rule cannot fully explain the DMI behavior here and one should seek for further physical mechanisms.

To explore these mechanisms, let us first analyze in details the impact of hydrogenation on electronic states of Co. In Fig. 2(a)-(d) we show the projected density of states (DOS) of Co atom in Co(1ML)@Gr, Co(1ML)@Gr-1/8H, Co(1ML)@Gr-2/8H, and Co(1ML)@Gr-4/8H. Due to the coverage of hydrogenated graphene on Co film, the Jahn-Teller distortion lowers the symmetry from D$_{3d}$ of pure Co film to C$_{3v}$ of Co(1ML)@Gr and Co(1ML)@Gr-4/8H and finally decreases to C$_s$ symmetry (i.e. no symmetry) of Co(1ML)@Gr-1/8 and Co(1ML)@Gr-2/8. Therefore, the five-fold degenerated 3$d$ electron states split into three irreducible representations, doublet degenerate level e$_1$ ($d_{yz}$, $d_{xz}$), e$_2$ ($d_{xy}$, $d_{x^2-y^2}$) and a singlet *a* state ($d_{z^2}$) in C$_{3v}$ and D$_{3d}$ point group. For C$_s$ symmetry, the e$_1$ and e$_2$ states are further broken as their corresponding orbitals split into non-degenerate states. For convenience, since the splitting of these states in C$_s$ symmetry has no dramatic influence on the DMI in our calculations, only $d_{xy}$, $d_{xz}$ and $d_{z^2}$ states for Co(1ML)@Gr-1/8H and Co(1ML)@Gr-2/8H systems are plotted in Fig. 2(b) and (c). Summarizing the variation of DOS with increase of H concentration in Fig. 2, one can note two characteristic features. First, bonding between Co and graphene gets stronger when H absorbed on graphene as $d$ states become more localized. Second, spin splitting decreases leading to smaller magnetic moments as seen in Fig. 1(d). As a result, the anti-bonding majority $d_{z^2}$ (minority $d_{xz}$) electronic states shift upward (downward) in energy with a decrease (increase) of their occupation. To better understand and visualize these findings on the influence of exchange splitting and crystal field on electronic states of Co, we plot the corresponding schematic diagram of the Co $d$ states energy levels in Co(1ML)@Gr and Co(1ML)/Gr-4/8H [Fig. 2(e) and (f)]. Note that e$_2$ ($d_{xy}$, $d_{x^2-y^2}$) states are spread across large region in energy (these energy levels are not shown). By comparing the two cases, one can see indeed that the majority $d_{z^2}$ states become less occupied since its anti-bonding state moves above the Fermi level (top brown line). Also, it clearly follows that the exchange splitting is smaller in case of Co(1ML)/Gr-4/8H.

In order to elucidate the origin of the sign and strength of DMI, we can now employ the analysis based on non-collinear Hamiltonian treated in the framework of the first-order perturbation theory [39-41]. The corrections to the total energy due to DMI can be approximated by the expectation value of the corresponding term written as $\langle\psi_{lm,s}|\xi\boldsymbol{\sigma L}|\psi_{lm,s}\rangle$, where $|\psi_{lm,s}\rangle = \sum|Y_{lm},\chi_s\rangle$ represents the quantum state comprising eigenstates $Y_{lm}$ and $\chi_s$ of orbital momentum and spin operators, respectively. Here $l, m$ and $s$ represent orbital, magnetic and spin quantum numbers, respectively. In our case, as shown in Fig. 1(a) and (b), the spins lie in $x$-$z$ plane, $\xi\sigma_x L_x$ and $\xi\sigma_z L_z$ components will cancel with each other, therefore, only the $y$ component $\langle\psi_{lm,s}|\xi\sigma_y L_y|\psi_{lm,s}\rangle$ needs to be considered [42]. Based on first-order perturbation theory, all of the occupied states contribute to the corrected total energy. For instance, following the analysis of SOC matrix elements between fully occupied different $d$ orbitals and spins, we find that the expectation value of $\langle d_{z^2}\chi_+ + d_{xz}\chi_-|\sigma_y L_y|d_{z^2}\chi_+ + d_{xz}\chi_-\rangle$ and $\langle d_{z^2}\chi_- + d_{xz}\chi_+|\sigma_y L_y|d_{z^2}\chi_- + d_{xz}\chi_+\rangle$ are respectively equal to $-\sqrt{6}$ and $\sqrt{6}$ giving the highest contribution to the DMI. In contrast, other non-zero matrix elements $\langle d_{xy}\chi_+ + d_{yz}\chi_-|\sigma_y L_y|d_{xy}\chi_+ + d_{yz}\chi_-\rangle$, $\langle d_{xz}\chi_+ + d_{x^2-y^2}\chi_-|\sigma_y L_y|d_{xz}\chi_+ + d_{x^2-y^2}\chi_-\rangle$,

$\langle d_{xy}\chi_- + d_{yz}\chi_+|\sigma_y L_y|d_{xy}\chi_- + d_{yz}\chi_+\rangle$ and $\langle d_{xz}\chi_- + d_{x^2-y^2}\chi_+|\sigma_y L_y|d_{xz}\chi_- + d_{x^2-y^2}\chi_+\rangle$ are equal to $\sqrt{2}, -\sqrt{2}, -\sqrt{2}$ and $\sqrt{2}$, respectively. Using the properties of rotation and Pauli spin matrices one can deduce that positive (negative) sign of matrix element of $\sigma_y L_y$ operator corresponds to the clockwise (anticlockwise) chirality of the DMI. Interestingly, similar analysis can be performed for magnetic anisotropy in the framework of second-order perturbation theory [43, 44].

The trend in DMI behavior as a function of hydrogen concentration can then be understood from the analysis of projected DOS for Co(1ML)@Gr shown in Fig. 2. One can see that only $d_{xz}$ minority and $d_z^2$ majority states are mostly affected by hydrogenation. Namely, increasing H concentration leads to increasing (decreasing) of $d_{xz}$ minority ($d_z^2$ majority) states occupation [Fig. 2(a)-(d)] so that among aforementioned matrix elements of SOC operator one should focus on the first one, i.e. $\langle d_{z^2}\chi_+ + d_{xz}\chi_-|\sigma_y L_y|d_{z^2}\chi_+ + d_{xz}\chi_-\rangle$, which corresponds to anticlockwise chirality. Since decreasing rate of $d_z^2$ majority occupation is higher than the increasing rate of $d_{xz}$ minority occupation, this SOC matrix element has an overall tendency to decrease. This causes lower DMI of ACW chirality for system without and with 1/8, 2/8 and 3/8 hydrogenation and eventual DMI change to CW chirality for Co(1ML)@Gr-4/8H [cf. Fig. 1(b) and Fig. 2(a)-(d)]. To further confirm this behavior, in Fig. 3 we plot the orbital resolved contributions to DMI deduced from SOC matrix elements for Co(1ML)@Gr and Co(1ML)@Gr-4/8H. It is clear that the strongest contribution to DMI comes indeed from the matrix element with $d_{xz}$ and $d_z^2$ orbitals, which is strongly negative and positive for Co(1ML)@Gr and Co(1ML)@Gr-4/8H as shown in Fig. 3(a) and (b), respectively.

To further verify the reliability of proposed fundamental mechanism responsible for strength and sign of DMI in hydrogenated Co@Gr structure, in Fig. 4 we present projected DOS of interfacial Co atom in structures with thicker Co, i.e. Co(7ML)@Gr and Co(7ML)@Gr-4/8H structures. Similar to structures with 1ML Co, two characteristic features, i.e. enlarged bonding between Co and carbon as well as the decreased spin splitting, are also observed in the 7ML Co structures. The DMI behavior is also similar giving anticlockwise and clockwise DMI for Co(7ML)@Gr and Co(7ML)@Gr-4/8H, respectively. Thus, we conclude that occupation of electronic states, notably $d_{xz}$ and $d_{z^2}$, are crucial for the strength and the sign of DMI, and their modulation is essential in controlling DMI.

In summary, using first principles calculations we have systematically investigated the effect of hydrogenation on DMI by using Co@Gr as a prototype system. We find that the DMI oscillates with a period of two ML as a function of Co film thickness. The DMI oscillations are mainly due to out-of-plane $d_{z^2(xz,yz)}$ orbitals. Furthermore, as hydrogen concentration increases, the anticlockwise DMI decreases and eventually changes its chirality to clockwise. This behavior can be explained in the framework of the first-order perturbation theory via analysis of occupation states of $d_{xz}$ and $d_{z^2}$ orbitals. This provides a promising method for controlling the spin-orbitronic phenomena such as the formation of domain walls and skyrmions. Moreover, our understanding of the DMI will serve as a guideline for experimental and theoretical investigations on the chiral magnetic systems.

## ACKNOWLEDGMENT

This work was supported by the National Natural Science Foundation of China (11874059), Zhejiang Province Natural Science Foundation of China (LR19A040002) and Ningbo 3315 project, Horizon 2020 Research and Innovation Programme under grant agreement no. 785219 (Graphene Flagship) and China Postdoctoral Science Foundation (2019M652151). The work was carried out at National Supercomputer Center in Tianjin, and the calculations were performed on TianHe-1 (A).

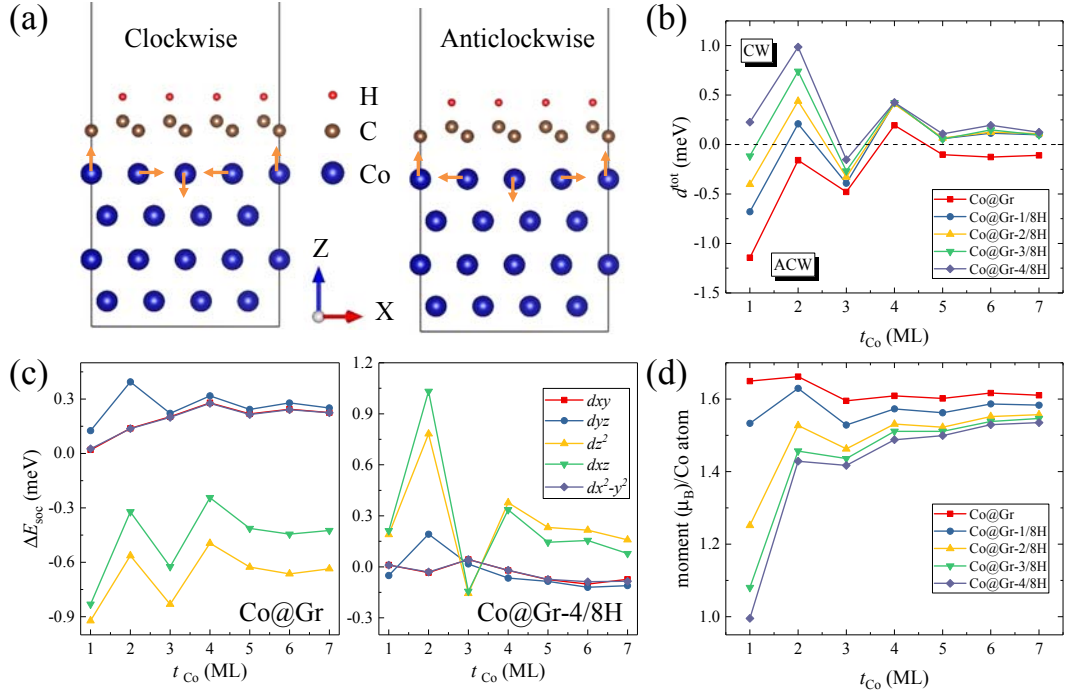

FIGURES.

FIG. 1. (a) Schematic diagram of Co(4ML)@Gr-4/8H with clockwise and anticlockwise spin textures. The arrows indicate spin orientations. (b) Calculated DMI as a function of Co thickness with different concentration of hydrogen on graphene surface. (c) SOC energy associated with DMI for Co@Gr and Co@Gr-4/8H as a function of Co thickness. (d) The average magnetic moment as a function of Co thickness with different concentration of hydrogen on graphene surface.

FIG. 2. (a-d) The density of states of Co in graphene/Co(1ML) systems with different concentration of hydrogen. The schematic diagram of energy levels of Co in (e) Co@Gr and (f) Co@Gr-4/8H.

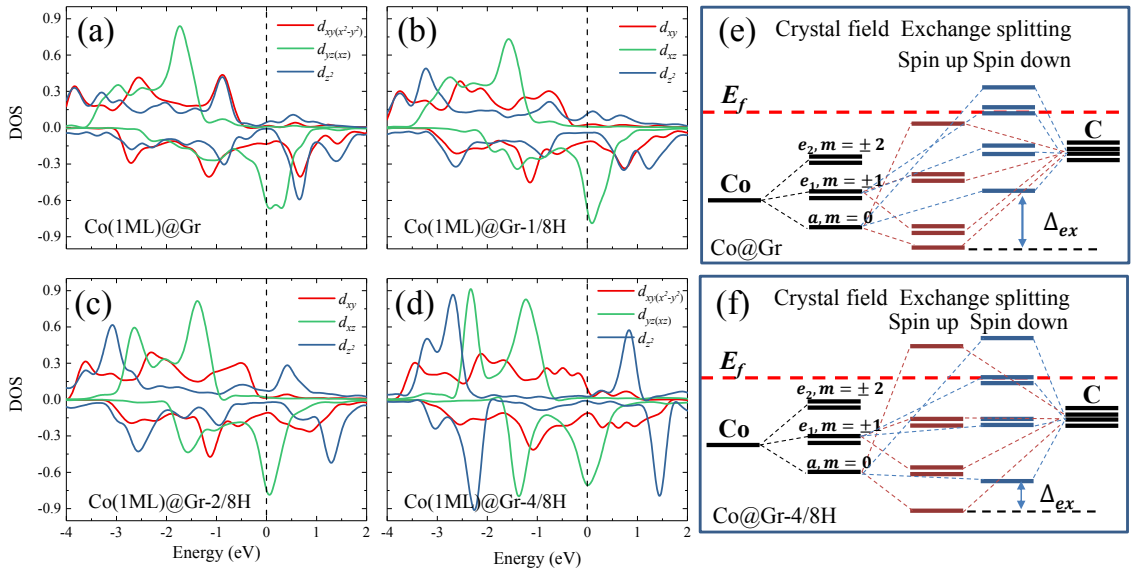

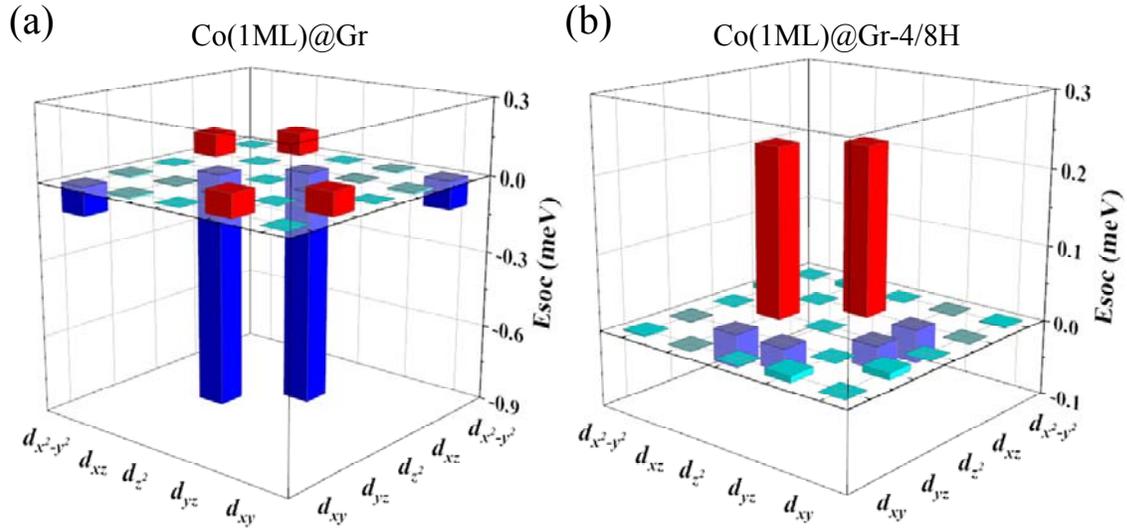

FIG. 3. Orbital resolved SOC energy matrix elements associated with DMI of Co in (a) Co(1ML)@Gr and (b) Co1(1ML)@Gr-4/8H, respectively.

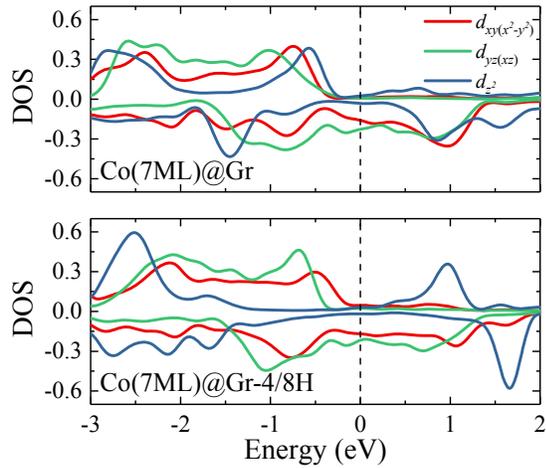

FIG. 4. The DOS of surface Co atom in (a) Co(7ML)@Gr and (b) Co(7ML)@Gr4/8H, respectively.